\pdfoutput=1
\documentclass{article}
\usepackage{arxiv}
\usepackage[utf8]{inputenc}
\usepackage[T1]{fontenc}
\usepackage{amsmath,amssymb}
\usepackage{graphicx}
\usepackage{booktabs}
\usepackage{array}
\usepackage{url}
\usepackage[round,semicolon,authoryear]{natbib}
\usepackage{hyperref}
\hypersetup{colorlinks=true, urlcolor=blue, linkcolor=blue, citecolor=blue}
\usepackage{doi}
\makeatletter
\def\@doi#1{%
  \let\#\relax \let\_\relax \let\textless\relax \let\textgreater\relax
  \edef\x{\toks0={{\doiurl#1}}}\x
  \edef\#{\@percentchar23}\edef\_{_}%
  \edef\textless{\@percentchar3C}\edef\textgreater{\@percentchar3E}%
  \edef\x{\toks2={\noexpand\href{\doiurl#1}}}\x
  \edef\x{\endgroup\the\toks2 \the\toks0}\x}
\makeatother
\usepackage{caption}
\DeclareUnicodeCharacter{0259}{e}
\DeclareUnicodeCharacter{0131}{\i}
\DeclareUnicodeCharacter{011F}{\u{g}}
\DeclareUnicodeCharacter{015F}{\c{s}}
\DeclareUnicodeCharacter{0130}{\.{I}}

\graphicspath{{figures/}}

\newcommand{\ZNFull}{ZN-Full}
\newcommand{\BCrisp}{B1}
\newcommand{\Bfloor}{\tilde{B}_{\mathrm{floor}}}
\newcommand{\Bhigh}{\tilde{B}_{\mathrm{high}}}

\title{An Eco-Friendly Tourism Trip Planner Driven by Z-Number Uncertainty: Evidence from Baku, Dubai, Abu Dhabi, Tashkent, and Istanbul}
\author{
  Aziz Nuriyev \\
  BA Programs \\
  Azerbaijan State Oil and Industry University \\
  Baku, Azerbaijan \\
  \texttt{aziz.nuriyev@asoiu.edu.az} \\
  \And
  Sherzod Turaev\thanks{Corresponding author.} \\
  College of Computing and Artificial Intelligence \\
  United Arab Emirates University \\
  Al Ain, United Arab Emirates \\
  \texttt{sherzod@uaeu.ac.ae} \\
  \AND
  Saja Al-Dabet \\
  Department of Artificial Intelligence \\
  Aqaba University of Technology \\
  Aqaba, Jordan \\
  \And
  Mary John \\
  Academic Support Department \\
  Abu Dhabi Polytechnic \\
  Abu Dhabi, United Arab Emirates \\
  \AND
  Kanan Qasimzade \\
  BA Programs \\
  Azerbaijan State Oil and Industry University \\
  Baku, Azerbaijan \\
  \And
  Alihuseyn Orujzada \\
  BA Programs \\
  Azerbaijan State Oil and Industry University \\
  Baku, Azerbaijan \\
  \AND
  Fatima Alblooshi \\
  College of Computing and Artificial Intelligence \\
  United Arab Emirates University \\
  Al Ain, United Arab Emirates \\
  \And
  Aisha Al Ali \\
  College of Computing and Artificial Intelligence \\
  United Arab Emirates University \\
  Al Ain, United Arab Emirates \\
}
\date{}

\begin{document}
\maketitle

\begin{abstract}
Deployed tourism trip planners ingest travel times and ratings as exact numbers, although routing services return travel times that vary with the hour of departure and a platform rating may rest on a handful of reviews or on hundreds of thousands. We present an eco-friendly trip planner that represents both quantities as \textit{Z-numbers}, which pair a fuzzy restriction on a value with a fuzzy reliability on the evidence behind it, and that constructs them directly from routing-service observations at six departure hours and from platform review counts rather than from linguistic elicitation. A hybrid loss combines score, carbon, travel-time, and reliability terms, and an adaptive large neighborhood search returns single-day itineraries through a conversational front end. We compare the planner with a crisp-mean baseline under an identical optimizer in Baku, Dubai, Abu Dhabi, Tashkent, and Istanbul, with fifteen points of interest and thirty seeds per city, and we evaluate the selected itineraries on their trip carbon and their total travel time. Pooled over 150 seed-paired runs, the planner lowers trip carbon from 3.24 to 2.61 kg (19.3\%, Wilcoxon $p = 0.017$; city-clustered bootstrap interval 6.4\% to 26.7\%) and travel time from 123 to 109 minutes per trip (11.5\%, $p = 2.8 \times 10^{-10}$; interval 5.0\% to 18.2\%). The travel-time reductions survive a Holm correction in Baku, Istanbul, and Abu Dhabi, the carbon reduction is significant in Baku and large but uncertain in the dispersed catalogs of Dubai and Abu Dhabi, and neither metric changes in Tashkent. Review counts are estimates of at least 3,000 for every point of interest, so rating reliability does not vary, and with a single baseline the results establish feasibility and direction rather than a general advantage. Code, routing caches, frozen routing tables, and per-seed results are openly released.
\end{abstract}


\section{Introduction}
\label{sec:introduction}

Digital technology has reshaped how travelers discover, evaluate, and sequence the attractions of a destination, and the itinerary that a tourist actually follows is now, to a considerable degree, the output of an information system rather than of a guidebook. The \textit{day-trip design problem}, in which a tourist arriving in an unfamiliar city must select a sequence of points of interest (POIs) and the transport options that connect them so as to balance experiential, monetary, temporal, and environmental criteria, has accordingly become one of the canonical applications of multi-criteria optimization \citep{gavalas2014survey,vansteenwegen2011survey}. Commercial planners such as TripAdvisor itineraries, Google Maps multi-stop routing, Sygic Travel, and Wanderlog, together with academic prototypes such as P-Tour \citep{maruyama2004ptour}, CT-Planner5 \citep{kurata2015ctplanner5}, and PersTour \citep{lim2018perstour}, address this task with considerable sophistication. What they do not address is the \textit{uncertainty} in the data on which every such planner relies.

That uncertainty has two sources. Routing services supply travel times that depend on the departure hour, the transport mode, and the recent traffic pattern, so that a single origin-destination-mode triple yields estimates that differ by a factor of two or more across the tourist-active day on congested corridors. Ratings carry uncertainty of a different kind: a POI with a mean of four and a half stars from five reviews carries less information than one with the same mean from five thousand, yet existing planners treat the two identically. A planner that ingests this evidence as crisp numbers discards the dimension of \textit{reliability} that the evidence itself supplies. The \textit{Z-number} formalism of \citet{zadeh2011znumber} represents an uncertain quantity as an ordered pair of fuzzy numbers, a \textit{restriction} on the value and a \textit{reliability} on the assertion, and so separates by construction the magnitude of an estimate from the sufficiency of the evidence on which it rests. Z-number arithmetic and ranking have been developed in depth \citep{aliev2016arithmetic,aliev2016ranking}, but Z-numbers have rarely served as the substrate of a working trip planner whose parameters come from operational data rather than from linguistic judgments.

In the current research, we build such a planner and test it. Its travel-time Z-numbers are constructed from Google Routes observations \citep{google2026routes} at six departure hours, its rating Z-numbers from the platform rating and review count of each POI, and its itineraries are optimized under a hybrid loss and returned through a conversational interface. The study pursues three objectives, which we state as research questions.
\begin{enumerate}
\item[\textbf{RQ1}] Can travel-time and rating Z-numbers be constructed from routing-service observations and review counts, without linguistic elicitation, and embedded in a working planner?
\item[\textbf{RQ2}] Under an identical optimizer, identical loss weights, and identical data, does the Z-number representation change the carbon footprint and the travel time of the selected itineraries relative to a crisp-mean representation of the same data?
\item[\textbf{RQ3}] How does any such effect vary across cities of contrasting geography, and which selection behavior accompanies it?
\end{enumerate}

We answer these questions on five cities of contrasting urban form, Baku, Dubai, Abu Dhabi, Tashkent, and Istanbul, and we compare the planner, denoted \ZNFull{}, with a crisp-mean baseline (\BCrisp{}) at thirty seeds per city. Because the two methods compute their loss terms by different formulas, we do not compare them on the loss; we evaluate the itineraries they select on physical outcomes measured identically for both, namely the trip carbon footprint, the total travel time, the transport modes used, and the across-seed stability of the selected POIs.

The contributions of the paper are three. \textit{First}, we specify data-driven construction operators that map routing observations and review counts to trapezoidal Z-numbers, a hybrid loss that combines them with crisp carbon and cost terms, and a two-step treatment of soft constraints, and we document a working implementation in sufficient detail for reproduction. \textit{Second}, we report a seed-paired comparison on five cities with per-city tests under a Holm correction and pooled tests supplemented by a city-clustered bootstrap, as a check that the pooled conclusions do not depend on treating seeds from the same city as independent, and we document a full re-run of the experiment from the released bundle. \textit{Third}, we characterize where the effect appears and where it does not, and we state plainly which parts of the evidence are measured and which are estimated. A companion study extends the comparison to five uncertainty representations on 349 POIs across the eight cities of the United Arab Emirates \citep{turaev2026evidence}; the present paper is the cross-national counterpart with a single baseline.

The paper is organized as follows. Section~\ref{sec:related-work} reviews the related work. Section~\ref{sec:framework} presents the Z-number framework, and Section~\ref{sec:system} the implemented system. Section~\ref{sec:case-study} specifies the case-study design, Section~\ref{sec:results} reports the results, Section~\ref{sec:discussion} discusses them together with the limitations, and Section~\ref{sec:conclusion} concludes.

\section{Related Work}
\label{sec:related-work}

\paragraph{Academic trip planners.} P-Tour \citep{maruyama2004ptour} allows a tourist to express preferences over time and POI categories and returns a sequenced itinerary that respects time windows. CT-Planner5 \citep{kurata2015ctplanner5} supports interactive tour planning that benefits both tourists and destinations, the team-orienteering variant is addressed by the greedy randomized adaptive search of \citet{souffriau2008grasp}, and PersTour \citep{lim2018perstour} personalizes POI recommendation from visit histories. Mobile and multimodal tour planners are surveyed and exemplified by \citet{gavalas2014survey} and \citet{gavalas2015ecompass}, and the orienteering problem on which all of these systems rest is surveyed by \citet{vansteenwegen2011survey}. Two features of this body of work motivate the present study. Travel times are treated as crisp scalars, drawn from a static matrix or from a single routing call per leg, so the variation that routing services expose across the day is not represented; and ratings are reduced to a single mean per POI, so a POI with abundant evidence and one with sparse evidence are scored identically when their means coincide.

\paragraph{Commercial planners.} TripAdvisor itineraries, Google Maps multi-stop routing, Sygic Travel, Visit-A-City, and Wanderlog operate at very large scale. Their user-facing output is crisp: a travel time is shown as a single number, and a rating as a single mean, with uncertainty acknowledged, if at all, through narrative hedges of the kind ``traffic may be heavy''. Their optimization layers are undocumented, so we treat them as a reference for the user experience rather than as methodological baselines.

\paragraph{Uncertainty in tourist routing.} Stochastic orienteering \citep{campbell2011stochastic,evers2014twostage} models travel and service times as random variables and optimizes an expected or chance-constrained objective, which requires a distributional assumption that a handful of routing observations does not support. Fuzzy formulations avoid that assumption: \citet{ruizmeza2021sustainable} develop a multi-objective fuzzy trip-design model with sustainable itineraries, \citet{exposito2019fuzzygrasp} propose a fuzzy GRASP for trip design with clustered POIs, and \citet{derya2024intuitionistic} combine an intuitionistic fuzzy POI score with exponentially distributed travel times. None of these representations, however, holds the spread of a value apart from the amount of evidence behind it. The Z-number formalism does \citep{zadeh2011znumber}, and its continuous arithmetic \citep{aliev2016arithmetic}, ranking \citep{aliev2016ranking}, and conversion to classical fuzzy numbers \citep{kang2012conversion} supply the operations a planner needs. To our knowledge, the Z-number literature on routing has relied on linguistically specified parameters, and the present work, together with its companion benchmark \citep{turaev2026evidence}, builds the parameters from operational data instead.

\section{The Z-Number Framework}
\label{sec:framework}

A \textit{Z-number} is an ordered pair $\tilde{Z} = (\tilde{A}, \tilde{B})$ of fuzzy numbers, in which $\tilde{A}$ is a restriction on the uncertain value and $\tilde{B} \subseteq [0,1]$ is the reliability with which the restriction is asserted \citep{zadeh2011znumber}. We work throughout with trapezoidal components $[a_1, a_2, a_3, a_4]$, $a_1 \le a_2 \le a_3 \le a_4$; a triangular component is the case $a_2 = a_3$, and a crisp number the case $a_1 = a_2 = a_3 = a_4$. Rating restrictions are triangular, reliabilities trapezoidal, and the monetary and carbon quantities crisp.

\subsection{The travel-time construction operator}
\label{sec:phi-T}

For an ordered origin-destination pair, the routing service is queried for driving and transit at the six departure hours 08:00, 10:30, 12:30, 14:30, 17:30, and 22:00, and it may return several routes per mode. The implementation identifies a route by its mode and its label, that is, the sequence of modes and transit lines it uses, so that the driving observations of a pair at all six hours form a single route and a transit route is a particular combination of lines. For a driving or transit route observed at the set of hours $\mathcal{H}$ with travel times $\tau_h$, $h \in \mathcal{H}$, the operator $\Phi_T$ assigns, for a departure at hour $t$, the restriction
\begin{equation}
\tilde{A}_T(t) = \bigl[\, \min_{h \in \mathcal{H}} \tau_h,\; \tau_{h^{*}(t)},\; \tau_{h^{*}(t)},\; \max_{h \in \mathcal{H}} \tau_h \,\bigr],
\label{eq:phi-T-A}
\end{equation}
in which $h^{*}(t)$ is the observed hour nearest to $t$ on the twenty-four-hour circle, so that the support spans the extremes observed across the day and the core is the observation closest in time to the actual departure; a route observed at a single hour receives a restriction of $\pm 10\%$ around that observation. The reliability is
\begin{equation}
\tilde{B}_T(N) = \Bfloor + \frac{N - 1}{N_{\max} - 1}\,\bigl(\Bhigh - \Bfloor\bigr),
\label{eq:phi-T-B}
\end{equation}
in which $N = |\mathcal{H}|$ is the number of departure hours at which the route was observed, $N_{\max} = 6$, $\Bfloor = [0.65, 0.70, 0.70, 0.80]$, and $\Bhigh = [0.85, 0.95, 0.95, 1.00]$. A waiting-time Z-number with the triangular restriction $(1, 5, 10)$ minutes is added to every driving and transit leg, and the crisp duration of such a leg, which the baseline uses and on which all reported travel times are based, includes the five-minute midpoint of this allowance. Walking legs, whose duration does not depend on traffic and which are queried at 12:30 only, receive the restriction $[0.85\,d, d, d, 1.15\,d]$ around their duration $d$ and the fixed reliability $[0.80, 0.90, 0.90, 1.00]$. The bounding trapezoids and the waiting allowance are design parameters of the implementation, fixed before the experiments reported here.

The restriction anchors its support at the observed extremes, so it inherits the empirical spread of a route across the day without a distributional assumption, while its core follows the departure time. The reliability reflects how consistently the route is available across the day: a route observed at all six hours receives $\Bhigh$, one observed at a single hour receives $\Bfloor$, and intermediate cases interpolate linearly. When observations are added, the support $[\min_h \tau_h, \max_h \tau_h]$ can only widen and $N$ can only grow, so the support of the restriction and the reliability are both monotone under set inclusion; the core, which is a single observation, is not.

\subsection{The rating construction operator}
\label{sec:phi-R}

For a POI with platform rating $\bar{s} \in [1, 5]$ and review count $n \ge 0$, the operator $\Phi_R$ assigns the restriction
\begin{equation}
\tilde{A}_R(\bar{s}) = \bigl[\, \max(1, 0.95\,\bar{s}),\; \bar{s},\; \bar{s},\; \min(5, 1.05\,\bar{s}) \,\bigr],
\label{eq:phi-R-A}
\end{equation}
divided by five times the number of POIs in the catalog, so that the summed restriction of any plan lies in the unit interval, and a reliability taken from a five-step ladder on the review count,
\begin{equation}
\tilde{B}_R(n) =
\begin{cases}
[0.85, 0.95, 0.95, 1.00], & n \ge 1{,}000,\\
[0.75, 0.85, 0.85, 0.95], & 500 \le n < 1{,}000,\\
[0.70, 0.78, 0.78, 0.88], & 100 \le n < 500,\\
[0.67, 0.74, 0.74, 0.82], & 20 \le n < 100,\\
[0.65, 0.70, 0.70, 0.80], & n < 20.
\end{cases}
\label{eq:phi-R-B}
\end{equation}
The restriction carries a fixed relative spread, and the evidence enters through the reliability alone, so that a POI with abundant evidence and one with sparse evidence are no longer scored identically when their means coincide. In the five catalogs studied here, however, every POI has at least 3,000 reviews and falls in the top step of the ladder, so the rating reliability is the same for all POIs and the rating channel does not discriminate by evidence in the experiments reported below (Section~\ref{sec:mechanism}).

\subsection{The hybrid loss functional}
\label{sec:loss}

The planner minimizes
\begin{equation}
\mathcal{L}(\pi) = w_s L_s + w_c L_c + w_t L_t + w_r L_r + \lambda_T V_T + \lambda_O V_O + \lambda_B V_B,
\label{eq:loss}
\end{equation}
in which $\pi$ is a candidate plan, that is, an ordered sequence of POIs together with the transport option chosen for each leg. The \textit{score} term $L_s$ is the Hellinger distance from the trip's aggregated rating Z-number to a reference Z-number whose restriction is the sum of the restrictions of all POIs in the catalog and whose reliability is maximal, $[1,1,1,1]$. The \textit{carbon} term $L_c$ is crisp: the trip's carbon footprint, at 0.192 kg of CO$_2$ per car kilometer and zero for walking and transit, divided by a fixed reference, the emission of driving at 50 km/h for the whole eight-hour window (76.8 kg), and capped at one. With trip footprints between 0 and about 12 kg in our runs, $w_c L_c$ never exceeds 0.031 and is typically about 0.006, so the carbon term contributes little to the loss (Section~\ref{sec:tradeoff}). The \textit{travel-time} term $L_t$ is the Hellinger distance from the sum of the travel-time Z-numbers of the plan's legs to the zero Z-number, normalized to the unit interval. The \textit{reliability} term $L_r$ is the mean, over the legs of the plan, of one minus the mean of the four parameters of the leg's travel-time reliability. The terms $V_T$, $V_O$, and $V_B$ penalize violations of the total-time, opening-hours, and budget soft constraints. In all experiments $w_s = 0.70$, $w_c = 0.20$, $w_t = 0.10$, $w_r = 0.10$, and $\lambda_T = \lambda_O = \lambda_B = 1$. The carbon and budget terms, which are the same in both methods, use ordinary arithmetic, since emission factors and prices are calibrated values rather than observed random quantities. The crisp baseline of Section~\ref{sec:case-study} replaces $L_s$ by one minus the ratio of the visited rating sum to an ideal sum and $L_t$ by the fraction of the time window spent traveling, and it has no reliability term; the loss values of the two methods are therefore not comparable, and we compare the itineraries instead.

\subsection{The two-step violation pattern}
\label{sec:two-step}

The total-time and opening-hours constraints compare Z-number aggregates with thresholds expressed as Z-numbers; the total-time threshold, for instance, has the restriction $(0.95\,T, T, 1.05\,T)$ for a window of $T$ minutes and a high reliability. We separate the \textit{decision} that a constraint is violated from the \textit{magnitude} of the violation. The decision is taken by a ranking comparison \citep{aliev2016ranking} in which the implementation gives the reliability a weight of zero, so that it rests on the restrictions alone and two aggregates whose restrictions agree are treated identically at this step whatever their reliabilities. Only when the decision is ``violated'' is a magnitude computed, as the Hellinger distance between the aggregate and the threshold, with the reliability reintroduced, normalized for the total-time constraint by the Hellinger distance of the threshold from zero and for the opening-hours constraints by that of a sixty-minute interval, in analogy with the relative excess of a crisp constraint. The penalty is therefore graded by the extent of the breach rather than imposed as an all-or-nothing indicator. The crisp baseline penalizes the relative excess of its crisp totals instead, so the pattern is part of the Z-number representation in the comparison below, and the experiments do not isolate its effect.

\subsection{The min-\texorpdfstring{$\tilde{B}$}{B} reliability rule}
\label{sec:min-B}

Exact reliability aggregation under Z-number addition \citep{aliev2016arithmetic} requires a family of optimization problems per aggregation, and in our implementation it costs between fifty and one hundred times as much as the rest of an optimization step. During search the planner therefore aggregates reliabilities by the \textit{min-$\tilde{B}$ rule}, which assigns to the reliability of a sum the componentwise minimum of the operands' reliabilities,
\begin{equation}
\tilde{B}_{1+2} = \bigl[\min(b_1^1, b_1^2),\, \min(b_2^1, b_2^2),\, \min(b_3^1, b_3^2),\, \min(b_4^1, b_4^2)\bigr].
\label{eq:min-B}
\end{equation}
The rule is a heuristic summary of the least reliable component of a plan, and it should not be read as a bound. In particular, it is not in general a lower bound on the reliability that the exact arithmetic assigns to the sum and can overstate it, since the evidence for a sum of uncertain quantities is in general weaker than the evidence for its weakest summand. The rule preserves the ordering of plans that differ in their least reliable leg, which is what the search requires of it, but the trip-level reliability it reports describes the weakest leg rather than certifying the plan. The experiments reported here do not compare the rule with the exact aggregation, and the reliabilities used in the loss are min-$\tilde{B}$ summaries.

\section{The Trip Planner System}
\label{sec:system}

The system is organized as a four-layer pipeline: data ingestion, Z-number construction, optimization, and user interface. The layers communicate through caches, so that ingestion runs on a slow, cost-aware schedule while the optimizer serves requests from cached data.

\paragraph{Data ingestion.} For the POI pairs of a city, the ingestion layer first queries the free OSRM table service for walking distances, which decide the pairs that receive options: DRIVE and TRANSIT for pairs at least 1 km apart and WALK for pairs within 3 km. It then queries the Google Routes API \citep{google2026routes} for DRIVE and TRANSIT at the six departure hours of Section~\ref{sec:phi-T}, in local time, and for WALK at 12:30, and it retains the duration, distance, cost, and emissions of the routes returned. Ratings, opening hours, dwell times, and entrance prices come from the city catalogs of Section~\ref{sec:case-study}, and review counts are estimates. Requests are rate-limited and retried with exponential backoff.

\paragraph{Z-number construction.} The routing observations are held in a versioned SQLite cache, and the construction layer applies $\Phi_T$ and $\Phi_R$ to them when a request is optimized.

\paragraph{Optimization.} Three interchangeable metaheuristics are available: local search with random restarts over swap, two-opt, option-swap, insert, and remove moves; adaptive large neighborhood search (ALNS) in the manner of \citet{ropke2006alns}, which is the default and the only optimizer used in the experiments; and a priority-encoded differential evolution \citep{storn1997de} for offline use.

\paragraph{User interface.} The planner is exposed as a FastMCP server to a conversational assistant, which collects the hotel base, the POIs of interest, the time window, the budget, and the allowed transport modes in a dialogue, and presents the three best plans returned by the server, each with the mode, duration, distance, cost, and emissions of every leg and the loss terms of the plan, together with a map. The loss weights, including the carbon weight $w_c$, are fields of the planner configuration, so that an experience-oriented or a sustainability-oriented setting can be chosen by changing one value.

\paragraph{Implementation and artifacts.} The system is written in Python with FastMCP, SQLite, NumPy, SciPy, and the \texttt{znum} library. The released bundle \citep{turaev2026zenodo} contains the source code under a CC BY 4.0 license, the five POI catalogs, the cached routing observations, the ingestion scripts, the seed-controlled driver, the frozen routing tables and run scripts of the re-run described in Section~\ref{sec:rerun}, and the per-seed itineraries, with per-leg modes and durations, from which every number in this paper is computed.

\section{The Case-Study Design}
\label{sec:case-study}

\subsection{Cities, catalogs, and scenario}

The five cities were chosen to cover contrasting urban forms: the compact center of Baku (Azerbaijan); the dispersed, transit-served layout of Dubai; the multi-node layout of Abu Dhabi, with its Corniche, Saadiyat Island, and Yas Island clusters (both United Arab Emirates); the compact center of Tashkent (Uzbekistan); and the dense historic peninsula of Istanbul (Turkey). Each catalog holds fifteen POIs within the municipal boundary, drawn from the city's tourism listings and covering cultural, leisure, commercial, and architectural attractions, with coordinates, rating, review count, opening hours, average dwell time, and entrance price. The catalog spreadsheets do not include review counts, so the counts used in all five cities are platform-consistent estimates; they range from 3,000 to 380,000, so every POI lies in the top step of the reliability ladder~(\ref{eq:phi-R-B}), a limitation to which we return in Section~\ref{sec:limitations}. Table~\ref{tab:catalog} lists the hotel bases, budgets, and taxi-fare parameters.

\begin{table}[!htb]
\centering
\caption{Per-city configuration. The hotel base is the origin and destination of every itinerary; the budget is the monetary soft constraint in local currency; the three taxi parameters define the linear taxi-cost model.}
\label{tab:catalog}
\small
\begin{tabular}{lllrrrrr}
\toprule
City & Hotel base & Currency & Budget & UTC & Taxi base & Per km & Per min \\
\midrule
Baku & Badamdar Hotel & AZN & 150 & $+4$ & 1.0 & 2.0 & 0.1 \\
Dubai & Address Downtown & AED & 500 & $+4$ & 12.0 & 1.96 & 0.5 \\
Abu Dhabi & Emirates Palace & AED & 500 & $+4$ & 12.0 & 1.82 & 0.5 \\
Tashkent & Hotel Uzbekistan & UZS & 750{,}000 & $+5$ & 5{,}000 & 3{,}500 & 500 \\
Istanbul & Four Seasons Sultanahmet & TRY & 5{,}000 & $+3$ & 27.5 & 15.0 & 5.0 \\
\bottomrule
\end{tabular}
\end{table}

The scenario is identical across cities: a single day starting at 10:00 local time with an eight-hour window, at most seven POIs, and the modes DRIVE, TRANSIT, and WALK. The loss weights and penalty coefficients of Section~\ref{sec:loss} are held fixed, so the cities differ only in geography, catalog, currency, and fares.

\subsection{Baseline}

The baseline \BCrisp{} uses the crisp travel time of each leg at the departure hour nearest to the actual departure, including the waiting allowance, and the platform rating of each POI, and it solves the resulting crisp problem with the same optimizer, weights, and penalty coefficients, so that the comparison isolates the Z-number representation as a whole. Because its loss terms are computed by different formulas (Section~\ref{sec:loss}), the two methods are compared only on the itineraries they return. We ran no other baselines; ablations that remove the reliability component alone, or that construct the Z-numbers linguistically, are left to future work, and a comparison against other uncertainty representations is reported in the companion benchmark \citep{turaev2026evidence}.

\subsection{Protocol and statistics}

Each method is run on each city at thirty seeds, with ALNS at three hundred iterations and three restarts per seed. In the re-run, a \ZNFull{} job took between about two and a half and seven minutes, depending on the machine (median 5.6 minutes), and a \BCrisp{} job 15 to 36 seconds. This yields 300 plans and 150 seed-paired comparisons. For each plan we record the trip carbon footprint in kilograms of CO$_2$ and the total travel time in minutes, both summed over the legs of the itinerary with the same emission and duration data for both methods (the travel time of a driving or transit leg including its five-minute waiting allowance), together with the transport mode of every leg. Per-city differences are tested with the two-sided Wilcoxon signed-rank test, and the ten per-city tests (five cities, two metrics) are corrected by the Holm procedure \citep{holm1979simple}; we also report Cohen's $d$ for the paired differences. Pooled tests over the 150 pairs treat seeds from the same city as independent, which overstates the effective sample size, so we supplement them with a city-clustered bootstrap that resamples the five cities with replacement (10,000 draws) and recomputes the pooled relative change. With only five clusters the bootstrap is coarse, and we read it as a check on the direction of the pooled effect rather than as a precise interval.

\subsection{Routing data and re-run}
\label{sec:rerun}

Table~\ref{tab:cache} reports the size of the released routing caches. All results in this paper come from a complete re-run of the 300 jobs from the released bundle, carried out in October 2026 with three changes that make the bundle reproducible without new data collection. \textit{First}, the cache expires entries after thirty days, and all entries were collected on 25 June 2026, so we fixed the cache clock at the date of the original runs. \textit{Second}, the released code queries the public OSRM server afresh on every run to decide which POI pairs receive walking, driving, and transit options, so we downloaded the OSRM tables once per city and used these frozen tables in every job; they are released with the bundle. \textit{Third}, the driver was extended to save the full itinerary, including the mode, duration, distance, and emissions of every leg. No new routing data were collected. Requests for pair-hours absent from the cache were sent without a valid key and rejected, so those options are unavailable to both methods: about one fifth of the drive and transit pair-hours requested in Tashkent and Abu Dhabi (576 of 2,808 and 528 of 2,880) and 4 of 2,557 in Istanbul. The exact reproduction of the original runs in Tashkent and Abu Dhabi indicates that the same options were unavailable then.

The re-run reproduces the originally reported per-seed itineraries exactly in Tashkent, Dubai, and Abu Dhabi (180 of 180 runs), and in 11 of the 120 runs in Baku and Istanbul, where the OSRM tables have changed since the original runs. The original runs gave a similar pooled carbon reduction (18.1\%, $p = 0.041$) and the same results in the three exactly reproduced cities, but in Baku the reduction was 14.5\% and in Istanbul the planner's carbon was 17.9\% higher than the baseline's, so the Istanbul carbon result is sensitive to the OSRM tables. The original runs did not record travel times in minutes, which is why the re-run is the basis of this paper.

\begin{table}[!htb]
\centering
\caption{Released routing caches: number of cached Routes API observations and file size per city.}
\label{tab:cache}
\small
\begin{tabular}{lrr}
\toprule
City & Observations & Cache size \\
\midrule
Baku & 7{,}145 & 17 MB \\
Dubai & 6{,}206 & 24 MB \\
Abu Dhabi & 5{,}708 & 22 MB \\
Tashkent & 2{,}325 & 4 MB \\
Istanbul & 5{,}026 & 12 MB \\
\midrule
Total & 26{,}410 & 79 MB \\
\bottomrule
\end{tabular}
\end{table}

\section{Results}
\label{sec:results}

\subsection{Per-city carbon and travel time}
\label{sec:per-city}

\begin{figure}[!htb]
\centering
\includegraphics[width=\textwidth]{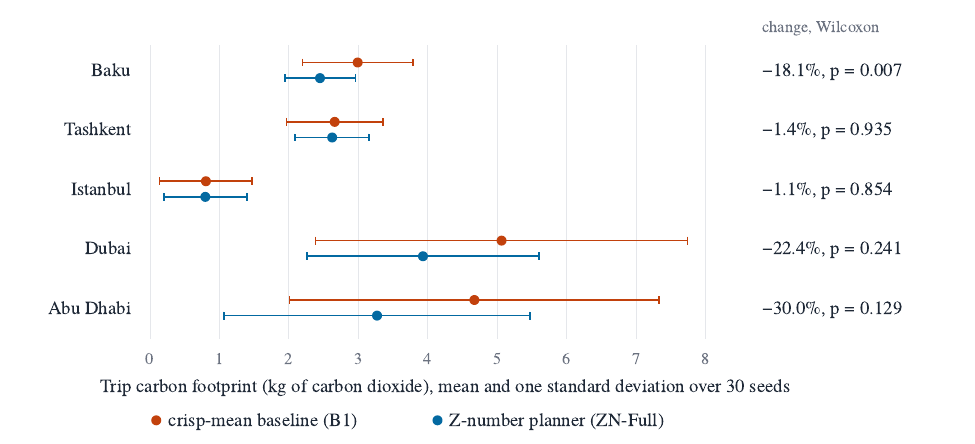}
\caption{Trip carbon footprint per city for the crisp-mean baseline (B1) and the Z-number planner (ZN-Full): mean and one standard deviation over thirty seeds, with the relative change and the two-sided seed-paired Wilcoxon $p$-value.}
\label{fig:carbon}
\end{figure}

Tables~\ref{tab:per-city} and~\ref{tab:per-city-tt} report the two outcomes per city, and Figure~\ref{fig:carbon} shows the carbon footprints.

\begin{table}[!htb]
\centering
\caption{Trip carbon footprint (kg of CO$_2$), mean $\pm$ standard deviation over thirty seeds. Change is the relative change of the \ZNFull{} mean against the \BCrisp{} mean; $p$ is the two-sided seed-paired Wilcoxon test; $p_{\mathrm{Holm}}$ is adjusted over the ten per-city tests of Tables~\ref{tab:per-city} and~\ref{tab:per-city-tt}; $d$ is Cohen's $d$ for the paired differences. $^{**}$ significant after the Holm correction.}
\label{tab:per-city}
\small
\begin{tabular}{lrrrrrr}
\toprule
City & \ZNFull{} & \BCrisp{} & Change & $p$ & $p_{\mathrm{Holm}}$ & $d$ \\
\midrule
Baku & $2.45 \pm 0.51$ & $2.99 \pm 0.80$ & $-18.1\%^{**}$ & 0.007 & 0.0498 & $-0.51$ \\
Tashkent & $2.62 \pm 0.53$ & $2.66 \pm 0.69$ & $-1.4\%$ & 0.935 & 1.000 & $-0.04$ \\
Istanbul & $0.80 \pm 0.60$ & $0.81 \pm 0.67$ & $-1.1\%$ & 0.854 & 1.000 & $-0.01$ \\
Dubai & $3.93 \pm 1.67$ & $5.06 \pm 2.68$ & $-22.4\%$ & 0.241 & 0.964 & $-0.31$ \\
Abu Dhabi & $3.27 \pm 2.21$ & $4.67 \pm 2.66$ & $-30.0\%$ & 0.129 & 0.647 & $-0.37$ \\
\bottomrule
\end{tabular}
\end{table}

Travel time responds more consistently than carbon. \ZNFull{} shortens the time spent traveling by 23.5\% in Abu Dhabi (28 minutes per trip), 15.2\% in Istanbul (18 minutes), and 10.7\% in Baku (15 minutes), and all three reductions survive the Holm correction, with large effect sizes in Baku and Abu Dhabi. Dubai shows a smaller reduction of 6.6\% that is not significant, and Tashkent is unchanged. On carbon, Baku shows a reduction of 18.1\%, about 0.5 kg per trip, which survives the correction at the margin ($p_{\mathrm{Holm}} = 0.0498$). Dubai and Abu Dhabi show the largest mean reductions, 22.4\% and 30.0\%, or 1.1 and 1.4 kg per trip, but the variance across seeds is high and neither is significant. Carbon is unchanged in Tashkent and in Istanbul, where both methods produce footprints below one kilogram.

\begin{table}[!htb]
\centering
\caption{Total travel time per trip (minutes), mean $\pm$ standard deviation over thirty seeds; columns as in Table~\ref{tab:per-city}.}
\label{tab:per-city-tt}
\small
\begin{tabular}{lrrrrrr}
\toprule
City & \ZNFull{} & \BCrisp{} & Change & $p$ & $p_{\mathrm{Holm}}$ & $d$ \\
\midrule
Baku & $125.5 \pm 10.6$ & $140.6 \pm 10.7$ & $-10.7\%^{**}$ & $1.8\times10^{-5}$ & $1.8\times10^{-4}$ & $-1.16$ \\
Tashkent & $84.5 \pm 5.8$ & $84.0 \pm 7.5$ & $+0.6\%$ & 0.773 & 1.000 & $+0.05$ \\
Istanbul & $98.3 \pm 16.9$ & $115.9 \pm 25.1$ & $-15.2\%^{**}$ & 0.001 & 0.010 & $-0.69$ \\
Dubai & $143.6 \pm 28.8$ & $153.7 \pm 25.2$ & $-6.6\%$ & 0.080 & 0.482 & $-0.28$ \\
Abu Dhabi & $92.4 \pm 28.8$ & $120.9 \pm 20.5$ & $-23.5\%^{**}$ & $2.4\times10^{-5}$ & $2.1\times10^{-4}$ & $-0.89$ \\
\bottomrule
\end{tabular}
\end{table}

In Abu Dhabi, the planner's converged plans cluster around Qasr Al Watan, Etihad Towers, Presidential Palace Park, and Corniche Beach, each selected on twenty-eight to thirty of the thirty seeds, while the baseline more often includes the Louvre Abu Dhabi (on 5 of 30 seeds against 1 for the planner), Al Bateen Beach (7 against 3), and Qasr Al Hosn (14 against 8).

\subsection{Pooled analysis}
\label{sec:pooled}

\begin{table}[!htb]
\centering
\caption{Pooled analysis over five cities and thirty seeds ($n = 150$ pairs). $p$ is the two-sided Wilcoxon test over the pooled pairs; the cluster interval resamples the five cities (10,000 draws) and is given for the relative change of the mean. Mode shares are shares of total travel minutes.}
\label{tab:pooled}
\footnotesize
\begin{tabular}{lrrrrr}
\toprule
Metric & \ZNFull{} & \BCrisp{} & Change & $p$ & Cluster 95\% \\
\midrule
Trip carbon (kg CO$_2$) & $2.61 \pm 1.66$ & $3.24 \pm 2.33$ & $-19.3\%$ & 0.017 & $[-26.7\%, -6.4\%]$ \\
Travel time (min per trip) & $108.9 \pm 30.0$ & $123.0 \pm 30.5$ & $-11.5\%$ & $2.8\times10^{-10}$ & $[-18.2\%, -5.0\%]$ \\
\midrule
POIs per plan & 6.06 & 6.24 & & & \\
Carbon per visited POI (kg) & 0.43 & 0.52 & & & \\
Travel minutes by car & 47.0\% & 45.0\% & & & \\
Travel minutes on foot & 26.6\% & 20.6\% & & & \\
Travel minutes by transit & 26.4\% & 34.3\% & & & \\
Strict core (POIs on 30/30) & 1.20 & 1.20 & & & \\
Loose core (POIs on $\ge$27/30) & 2.60 & 2.80 & & & \\
\bottomrule
\end{tabular}
\end{table}

\begin{figure}[!htb]
\centering
\includegraphics[width=\textwidth]{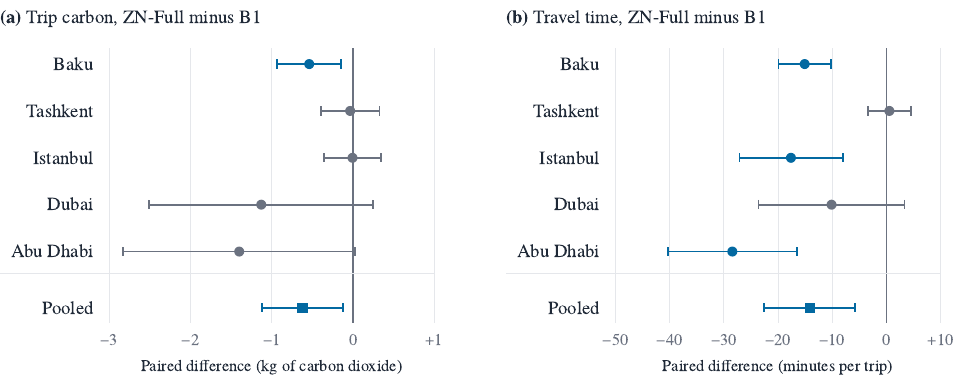}
\caption{Paired differences, ZN-Full minus B1, for (a) trip carbon and (b) travel time per trip. Circles are per-city mean differences over thirty seeds with 95\% $t$-intervals; squares are the pooled mean difference with a 95\% city-clustered bootstrap interval. Intervals that exclude zero on the favorable side are drawn in blue.}
\label{fig:paired}
\end{figure}

Table~\ref{tab:pooled} and Figure~\ref{fig:paired} summarize the pooled comparison. Pooled over the 150 pairs, \ZNFull{} lowers mean trip carbon from 3.24 to 2.61 kg, a reduction of 19.3\% ($p = 0.017$), and mean travel time from 123.0 to 108.9 minutes, a reduction of 11.5\% ($p = 2.8 \times 10^{-10}$). Both reductions keep their sign under the city-clustered bootstrap, whose intervals exclude zero. The mean favors \ZNFull{} in four of the five cities on travel time and is lower in all five on carbon, although the Tashkent and Istanbul carbon differences are negligible and the Istanbul sign depends on the routing tables (Section~\ref{sec:rerun}). Plans of the two methods are of nearly equal length (6.06 against 6.24 POIs), so the reductions do not come mainly from shorter itineraries; carbon per visited POI is lower for \ZNFull{} in four of the five cities, the exception being Istanbul, where both are near 0.12 kg. The planner's itineraries spend a larger share of their travel time on foot (26.6\% against 20.6\%) and a smaller share on transit (26.4\% against 34.3\%), while the share by car is similar.

\subsection{Selection stability and constraints}
\label{sec:convergence}

The strict across-seed core, the POIs selected on all thirty seeds, averages 1.20 per city for both methods, and under the looser threshold of at least 27 of 30 seeds the baseline's core is slightly larger (2.80 against 2.60). The two methods are therefore similarly stable, and we draw no conclusion from selection stability. Soft-constraint violations are rare for both, occurring on 7 of the 150 \ZNFull{} runs and 9 of the 150 \BCrisp{} runs.

\section{Discussion}
\label{sec:discussion}

\subsection{Where the effect appears}

The effect varies across cities in a way that is consistent with the geographic dispersion of their catalogs, although five cities cannot establish such a relationship and we did not measure dispersion quantitatively. In the dispersed catalogs of Dubai and Abu Dhabi, the planner's mean carbon is 1.1 to 1.4 kg lower per trip and, in Abu Dhabi, its travel time is almost half an hour shorter; dispersion creates opportunities for avoiding distant POIs and for mode substitution, and it is here that the representation changes the outcome most. In the compact center of Baku and the walkable historic center of Istanbul, the planner shortens travel by a quarter of an hour or more, and in Baku this also lowers carbon, whereas in Istanbul the footprints of both methods are already below one kilogram. In Tashkent the two methods select different POI sets on every seed but reach nearly the same carbon and travel time; we have no explanation for this from the present data, and the missing drive and transit options cannot account for it, since Abu Dhabi lacks a similar share and shows the largest effect.

\subsection{The mechanism}
\label{sec:mechanism}

The comparison isolates the representation as a whole, so it cannot apportion the effect among the restriction spread, the reliability component, and the two-step violation pattern. Two observations constrain the interpretation. First, the travel-time term of \ZNFull{} measures a plan through Z-numbers whose support spans the extremes observed across the day, not only the duration at the departure hour, so legs whose duration varies strongly across the day are penalized more than in the baseline, which favors itineraries with short legs of small absolute spread; the larger share of walking time in the planner's itineraries is consistent with this. Second, in Abu Dhabi the baseline more often admits distant POIs whose crisp scores are competitive, while the planner retains a compact set near the hotel base. The rating channel, by contrast, does not discriminate by evidence in these catalogs: every POI falls in the top step of the reliability ladder, and the rating restriction has the same relative spread for every POI apart from the clamp at the rating ceiling, so the rating Z-numbers differ from the crisp ratings only through the geometry of the Hellinger distance, not through their evidence. The effect reported here is therefore most plausibly carried by the travel-time channel, although an ablation would be needed to establish this.

The companion benchmark across the eight cities of the United Arab Emirates, which compares five representations on 349 POIs under the same planner, reaches a compatible but more cautious conclusion on carbon \citep{turaev2026evidence}: representations with an explicit evidence channel admit no poorly evidenced POIs, and the Z-number planner's lower trip carbon there arises largely from scheduling fewer POIs, with a per-POI carbon that is not robustly lower than that of the crisp planner. In the present study the plans are of similar length and the per-POI carbon is lower, so the two studies differ in how the carbon reduction arises, which is itself a reason not to generalize from either.

\subsection{The carbon weight}
\label{sec:tradeoff}

The default weights give the score term three and a half times the weight of the carbon term, and with the fixed reference of 76.8 kg the weighted carbon term is at most 0.031 and typically two to three percent of the loss, so the carbon term is nearly inert in both methods; the planner nonetheless does not raise mean carbon in any of the five cities, which suggests that its carbon reductions come from the representation of travel time rather than from pricing emissions. We did not sweep the carbon weight, so the trade-off between score and carbon along that weight is not characterized here. Because $w_c$ is a single configuration value, such a sweep, a reference emission scaled to the trip, and a study of the settings that users actually choose are straightforward follow-ups.

\subsection{Limitations}
\label{sec:limitations}

Seven limitations scope the findings. \textit{First}, review counts are estimates for all seventy-five POIs and are all at least 3,000, so the rating reliability is the same for every POI and the evidence channel on ratings is not exercised by these catalogs. \textit{Second}, the comparison involves a single crisp-mean baseline and no ablation, so it shows that the Z-number representation as a whole changes the selected plans but not which of its components is responsible. \textit{Third}, each catalog holds fifteen POIs, and larger catalogs, which the system supports, may behave differently. \textit{Fourth}, only one scenario is reported, a ten o'clock start with an eight-hour window and at most seven POIs. \textit{Fifth}, the pooled inference rests on five cities; the city-clustered bootstrap confirms the sign of the pooled effects but is coarse with so few clusters. \textit{Sixth}, the routing data are a single snapshot, the cache lacks a fifth of the drive and transit options requested in Tashkent and Abu Dhabi, and the selection of pairs that receive each mode rests on OSRM tables frozen in October 2026, which differ from those of the original runs in Baku and Istanbul. \textit{Seventh}, the evaluation is computational: we did not study whether tourists understand, trust, or follow reliability information, which is the question on which the practical value of an evidence-aware planner ultimately depends. The min-$\tilde{B}$ reliability summaries, finally, are heuristic rather than guaranteed (Section~\ref{sec:min-B}).

\section{Conclusion}
\label{sec:conclusion}

We have presented a tourism trip planner that represents travel times and ratings as Z-numbers constructed from routing-service observations and review counts, optimizes itineraries under a hybrid loss with a carbon term, and returns them through a conversational interface, at a cost of a few minutes of computation per request at the reported search budget (RQ1). Against a crisp-mean baseline under an identical optimizer in five cities, the planner lowers pooled trip carbon by 19.3\% and pooled travel time by 11.5\%, with both effects keeping their sign under a city-clustered bootstrap (RQ2). Travel time falls significantly in Baku, Istanbul, and Abu Dhabi; carbon falls significantly in Baku and by the largest amounts, though with high variance, in the dispersed catalogs of Dubai and Abu Dhabi; and neither changes in Tashkent (RQ3). With a single baseline, estimated and uniformly high review counts, and fifteen POIs per city, the study establishes that a data-driven Z-number planner is feasible and that its effect on carbon and travel time is favorable in direction and varies across cities, rather than that it outperforms crisp planning in general. The re-run from the released bundle reproduces the original itineraries exactly in three of the five cities, and the remaining differences, which change the sign of a negligible carbon difference in Istanbul, show how sensitive such comparisons are to the routing data. Ablations of its components, larger catalogs with measured and sparse review counts, a sweep of the carbon weight, and a study of how tourists respond to reliability information are the natural next steps.

\section*{Funding}
This work was supported by the United Arab Emirates University through the UAEU-ASOIU Collaborative SDG Research Program Grant (grant number G00005329), administered via the Emirates Center for Mobility Research. The funder had no role in the study design, the collection or interpretation of the data, the writing of the report, or the decision to submit it for publication.

\section*{Competing interests}
The authors declare no competing interests.

\section*{Data and code availability}
The source code, the five POI catalogs, the cached Routes API observations, the ingestion scripts, the seed-controlled driver, the frozen OSRM tables and scripts of the re-run, and the per-seed itineraries underlying every table and figure are openly available on Zenodo at \url{https://doi.org/10.5281/zenodo.23522063} under a CC BY 4.0 license \citep{turaev2026zenodo}.

\section*{Author contributions}
\textbf{Aziz Nuriyev:} Conceptualization, Methodology, Formal analysis, Investigation, Validation, Writing (original draft). \textbf{Sherzod Turaev:} Conceptualization, Methodology, Formal analysis, Investigation, Supervision, Funding acquisition, Writing (original draft), Writing (review and editing), Project administration. \textbf{Saja Al-Dabet:} Investigation, Validation, Writing (review and editing). \textbf{Mary John:} Investigation, Validation, Writing (review and editing). \textbf{Kanan Qasimzade:} Software, Data curation, Investigation. \textbf{Alihuseyn Orujzada:} Software, Data curation, Investigation. \textbf{Fatima Alblooshi:} Investigation, Validation. \textbf{Aisha Al Ali:} Investigation, Validation.

\setlength{\bibsep}{2pt plus 1pt}
\bibliographystyle{plainnat-doi}
\bibliography{references}

@article{lim2018perstour,
  author  = {Lim, K.~H. and Chan, J. and Leckie, C. and Karunasekera, S.},
  title   = {Personalized trip recommendation for tourists based on user interests, points of interest visit durations and visit recency},
  journal = {Knowledge and Information Systems},
  volume  = {54},
  number  = {2},
  pages   = {375--406},
  year    = {2018},
  doi     = {10.1007/s10115-017-1056-y}
}

@inproceedings{maruyama2004ptour,
  author    = {Maruyama, A. and Shibata, N. and Murata, Y. and Yasumoto, K. and Ito, M.},
  title     = {{P-TOUR}: A personal navigation system for tourism},
  booktitle = {Proc. 11th World Congress on Intelligent Transport Systems},
  pages     = {18--21},
  year      = {2004}
}

@inproceedings{kurata2015ctplanner5,
  author    = {Kurata, Y. and Shinagawa, Y. and Hara, T.},
  title     = {{CT-Planner5}: A computer-aided tour planning service which profits both tourists and destinations},
  booktitle = {Workshop on Tourism Recommender Systems ({TouRS'15}) at {RecSys'15}},
  address   = {Vienna, Austria},
  month     = sep,
  year      = {2015}
}

@article{gavalas2014survey,
  author  = {Gavalas, D. and Konstantopoulos, C. and Mastakas, K. and Pantziou, G.},
  title   = {A survey on algorithmic approaches for solving tourist trip design problems},
  journal = {Journal of Heuristics},
  volume  = {20},
  number  = {3},
  pages   = {291--328},
  year    = {2014},
  doi     = {10.1007/s10732-014-9242-5}
}

@article{gavalas2015ecompass,
  author  = {Gavalas, D. and Kasapakis, V. and Konstantopoulos, C. and Pantziou, G. and Vathis, N. and Zaroliagis, C.},
  title   = {The {eCOMPASS} multimodal tourist tour planner},
  journal = {Expert Systems with Applications},
  volume  = {42},
  number  = {21},
  pages   = {7303--7316},
  year    = {2015},
  doi     = {10.1016/j.eswa.2015.05.046}
}

@article{vansteenwegen2011survey,
  author  = {Vansteenwegen, P. and Souffriau, W. and Van Oudheusden, D.},
  title   = {The orienteering problem: A survey},
  journal = {European Journal of Operational Research},
  volume  = {209},
  number  = {1},
  pages   = {1--10},
  year    = {2011},
  doi     = {10.1016/j.ejor.2010.03.045}
}

@inproceedings{souffriau2008grasp,
  author    = {Souffriau, W. and Vansteenwegen, P. and Vanden Berghe, G. and Van Oudheusden, D.},
  title     = {A greedy randomised adaptive search procedure for the team orienteering problem},
  booktitle = {EU/MEeting 2008 Workshop on Metaheuristics for Logistics and Vehicle Routing},
  year      = {2008}
}

@article{campbell2011stochastic,
  author  = {Campbell, A.~M. and Gendreau, M. and Thomas, B.~W.},
  title   = {The orienteering problem with stochastic travel and service times},
  journal = {Annals of Operations Research},
  volume  = {186},
  number  = {1},
  pages   = {61--81},
  year    = {2011},
  doi     = {10.1007/s10479-011-0895-2}
}

@article{evers2014twostage,
  author  = {Evers, L. and Glorie, K. and van der Ster, S. and Barros, A.~I. and Monsuur, H.},
  title   = {A two-stage approach to the orienteering problem with stochastic weights},
  journal = {Computers and Operations Research},
  volume  = {43},
  pages   = {248--260},
  year    = {2014},
  doi     = {10.1016/j.cor.2013.09.011}
}

@article{exposito2019fuzzygrasp,
  author  = {Exp\'osito, A. and Mancini, S. and Brito, J. and Moreno, J.~A.},
  title   = {A fuzzy {GRASP} for the tourist trip design with clustered {POIs}},
  journal = {Expert Systems with Applications},
  volume  = {127},
  pages   = {210--227},
  year    = {2019},
  doi     = {10.1016/j.eswa.2019.03.004}
}

@article{ruizmeza2021sustainable,
  author  = {Ruiz-Meza, J. and Brito, J. and Montoya-Torres, J.~R.},
  title   = {Multi-objective fuzzy tourist trip design problem with heterogeneous preferences and sustainable itineraries},
  journal = {Sustainability},
  volume  = {13},
  number  = {17},
  pages   = {9771},
  year    = {2021},
  doi     = {10.3390/su13179771}
}

@article{derya2024intuitionistic,
  author  = {Derya, T. and Atalay, K. D. and Dinler, E. and Ke\c{c}eci, B.},
  title   = {Selective clustered tourist trip design problem with time windows under intuitionistic fuzzy score and exponential travel times},
  journal = {Expert Systems with Applications},
  volume  = {255},
  pages   = {124792},
  year    = {2024},
  doi     = {10.1016/j.eswa.2024.124792}
}

@article{zadeh2011znumber,
  author  = {Zadeh, L.~A.},
  title   = {A note on {Z}-numbers},
  journal = {Information Sciences},
  volume  = {181},
  number  = {14},
  pages   = {2923--2932},
  year    = {2011},
  doi     = {10.1016/j.ins.2011.02.022}
}

@article{kang2012conversion,
  author  = {Kang, B. and Wei, D. and Li, Y. and Deng, Y.},
  title   = {A method of converting {Z}-number to classical fuzzy number},
  journal = {Journal of Information and Computational Science},
  volume  = {9},
  number  = {3},
  pages   = {703--709},
  year    = {2012}
}

@article{aliev2016arithmetic,
  author  = {Aliev, R.~A. and Huseynov, O.~H. and Zeinalova, L.~M.},
  title   = {The arithmetic of continuous {Z}-numbers},
  journal = {Information Sciences},
  volume  = {373},
  pages   = {441--460},
  year    = {2016},
  doi     = {10.1016/j.ins.2016.08.078}
}

@article{aliev2016ranking,
  author  = {Aliev, R.~A. and Huseynov, O.~H. and Serdaroglu, R.},
  title   = {Ranking of {Z}-numbers and its application in decision making},
  journal = {International Journal of Information Technology and Decision Making},
  volume  = {15},
  number  = {6},
  pages   = {1503--1519},
  year    = {2016},
  doi     = {10.1142/S0219622016500310}
}

@article{ropke2006alns,
  author  = {Ropke, S. and Pisinger, D.},
  title   = {An adaptive large neighborhood search heuristic for the pickup and delivery problem with time windows},
  journal = {Transportation Science},
  volume  = {40},
  number  = {4},
  pages   = {455--472},
  year    = {2006},
  doi     = {10.1287/trsc.1050.0135}
}

@article{storn1997de,
  author  = {Storn, R. and Price, K.},
  title   = {Differential evolution: A simple and efficient heuristic for global optimization over continuous spaces},
  journal = {Journal of Global Optimization},
  volume  = {11},
  number  = {4},
  pages   = {341--359},
  year    = {1997},
  doi     = {10.1023/A:1008202821328}
}

@article{holm1979simple,
  author  = {Holm, S.},
  title   = {A simple sequentially rejective multiple test procedure},
  journal = {Scandinavian Journal of Statistics},
  volume  = {6},
  number  = {2},
  pages   = {65--70},
  year    = {1979}
}

@misc{google2026routes,
  author       = {{Google}},
  title        = {Routes {API} documentation},
  howpublished = {Google Maps Platform Developer Documentation},
  year         = {2026},
  note         = {Accessed 2026. [Online]. Available: \url{https://developers.google.com/maps/documentation/routes}}
}

@misc{turaev2026zenodo,
  author       = {Nuriyev, Aziz and Turaev, Sherzod and Al-Dabet, Saja and John, Mary},
  title        = {Cross-city {Z}-number trip planner: Reproduction bundle with routing caches, frozen routing tables, re-run itineraries, and analysis scripts},
  year         = {2026},
  publisher    = {Zenodo},
  howpublished = {Dataset},
  doi          = {10.5281/zenodo.23522063}
}

@article{turaev2026evidence,
  author  = {Turaev, Sherzod and John, Mary and Nuriyev, Aziz and Al-Dabet, Saja},
  title   = {Evidence-aware uncertainty representation for eco-friendly tourism trip planning: A national benchmark of {Z}-numbers against alternative formalisms across the {United Arab Emirates}},
  journal = {Complex \& Intelligent Systems},
  year    = {2026},
  note    = {Accepted, in press}
}

\end{document}